\documentclass[aps,twocolumn,reprint]{revtex4}
\makeatletter
\def\@dotsep{4.5}
\makeatother

\usepackage{graphicx}
\usepackage{amssymb}
\usepackage{natbib}
\usepackage{amsmath}

\newcommand{\comment}[1]{}

\begin{document}

\title{Insights into the flux of water in a water desalination  through nanopores process}

\author{Cl\'audia K. B. de Vasconcelos$^{1,2}$}
\author{Ronaldo J. C. Batista$^{3}$}
\author {McGlennon da Rocha R\'egis$^{4}$}
\author{Ta\'{i}se M. Manhabosco$^{3}$}
\author{Alan B. de Oliveira$^{3}$}

\affiliation{$^{1}$ Escola de Minas, Universidade Federal de Ouro Preto, Ouro Preto, MG 35400-000, Brazil.\\
$^{2}$ Departamento de F\'{i}sica e Qu\'{i}mica, PUC Minas, Belo Horizonte, MG 30535-901, Brazil.\\
$^{3}$ Departamento de F\'{i}sica, Universidade Federal de Ouro Preto, Ouro Preto, MG 35400-000, Brazil.\\
$^{4}$ Instituto Federal de Minas Gerais -- Campus Congonhas, Congonhas, MG 36415-000, Brazil.}

\begin{abstract}

Water desalination through nanopores has been shown to be a promising alternative to 
the currently water purification processes. In spite the results in this direction obtained by means of computational simulations were animating there are still pending issues  to be resolved. For example, water desalination 
involves macro numbers (in size and time) but  in such a scale it is literally impossible to attack this problem
using all-atoms  simulations. It is common to extrapolate results from nano to macro
sizes in order to estimate quantities of interest, which must be taken with care. Here we present 
a simple model which mimics the separation of salt from water, which may help
to attack bigger problems on water desalination subjects. Besides, we show that the investigation of a restrict space of parameters imposed by
expensive models may hidden interesting, important features involved in the water desalination problem. Finally, we present an analytical calculation which explains the rich behaviour of the  water flux    through nanopores in a water-salt separation scenario.

\end{abstract}

\maketitle


The shortage of potable water  has been an increasingly relevant problem. Processes for obtaining clean water are still inefficient and tend to be prohibitively expensive. The seawater desalination seems to be a promising alternative: despite approximately 97\% of the water is concentrated in the oceans and seas the percentage of potable water obtained through this means is still very small \cite{elimelech2011future}. Unfortunately all known solute-solvent  separation  based processes  are   also very inefficient. Examples are distillation, reverse osmosis, thermal desalination and freezing \cite{elimelech2011future, spiegler2001energetics,fornasiero2008ion}.

Mechanisms of water desalination based on nanostructures have obtained encouraging results \cite{bae2010roll}.  The size exclusion promoted by nanopores 
is a key ingredient when such structures are used as filters. For example, Cohen-Tanugi and Grossman have recently reported  results from simulations of water desalination through a graphene membrane.
Their findings  indicate that nanostructures are capable of rejecting salt ions and increase the water flow and permeability in several orders of magnitude if compared with existing processes of water-salt separation \cite{cohen2012water}. Recently the water-ions separation trough nanosctructures subject gained
 attention of the scientific community, who have intensified the efforts in order
to understand the main mechanisms responsible for water desalination in such a nanometric level 
\cite{das2014carbon,kou2013water,rish2014exceptional,liu2014mechanical,
cohen2014water,xue2013exceptionally,ramallo2013effective,yang2014sponge,
nicolai2014tunable,hu2013enabling,sun2014mechanisms,
konatham2013simulation,thomas2014have}.

One problem when approaching all-atoms models is that they are generally computationally expensive, forcing the investigation to be restricted to small systems, short times, and to a modest variety of parameters.  Since the desalination problem is typically macro any extrapolation to real world quantities must be performed with care. In this sense, we need either more powerful machines or to build cheaper models. In this work we present a simple model  for water-salt separation having core-softened potentials as a building block. We show that it is possible to build a system which qualitatively reproduces the results
of more complexes simulations in a fraction of time. Also, we present a theoretical model which explains the water flux  through nanopores behaviour for a wide variety of parameters. We believe this study may help to understand the water desalination through nanostructures process in a more realistic number of particles  and parameters scenario.

\begin{figure}[ht]
\centering
\includegraphics[width=0.35\textwidth]{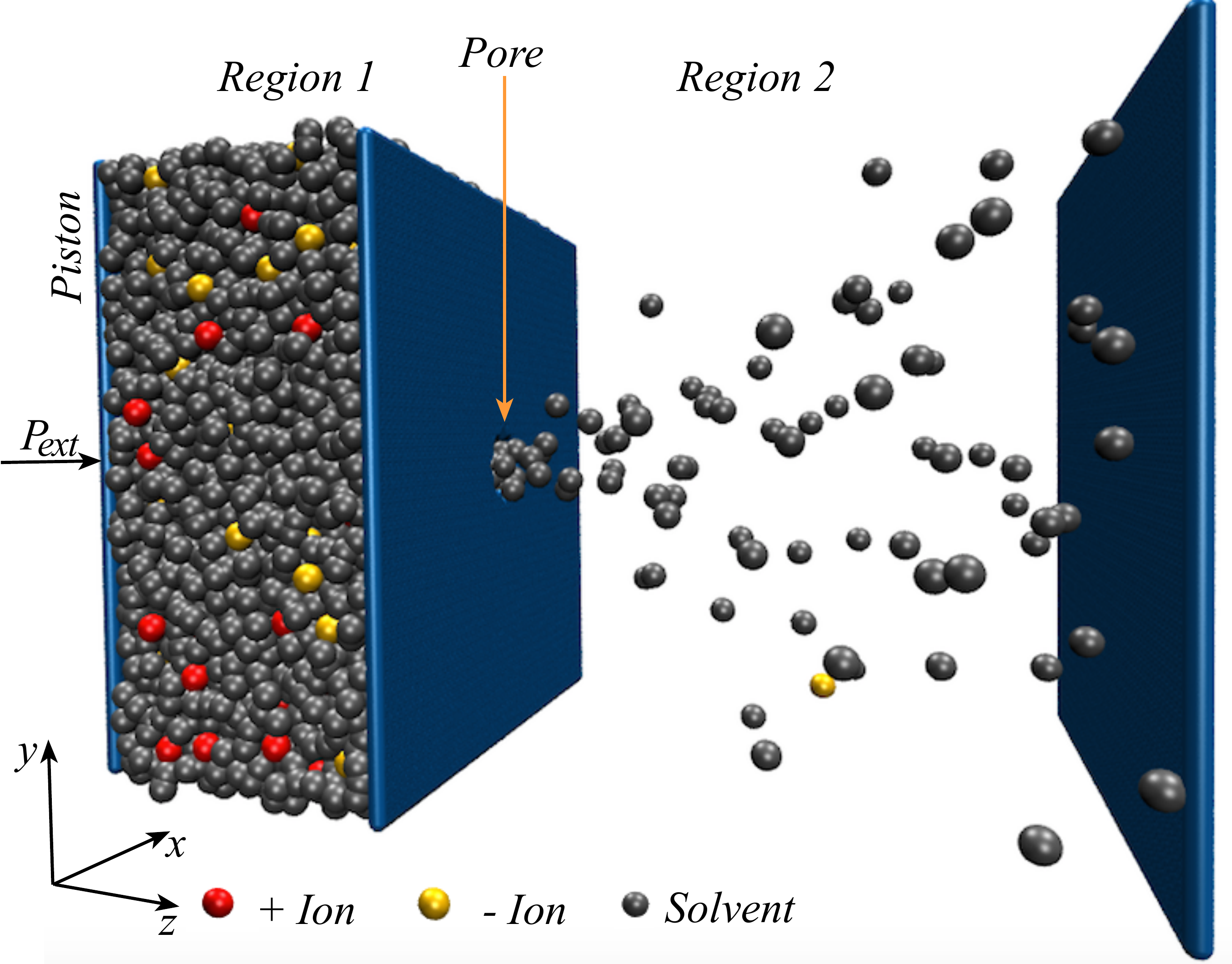}
\caption{System model we have used for mimicking the water desalination through nanopores process. The mixture of solvent and ions, initially localised in the Region 1, is pushed by a piston against a middle membrane. The circular pore in its center prioritises the passage of solvent once ions are effectively bigger than solvent particles.}
\label{system}
\end{figure}

Our system  is shown in Figure \ref{system}. We used a simulation box with dimensions of 50$\times$50$\times$75, in units of  size of solvent particles $\sigma$, in $x$, $y$ and $z,$ directions, respectively. There is a piston initially located at $z=0$, in which we apply an external pressure, and two walls. The middle wall, whose position is fixed at $z=25\sigma$,  separates the Region 1 from Region 2 (see Fig. \ref{system}) and it has a circular porous in its center. Finally, there is the end wall, fixed at $z=75\sigma$, closing the system in the $z$ direction. Periodic boundary conditions were used in the other directions. Piston and walls were constituted by 2500 ``frozen'' particles, equally distributed in a 
single layer. In the specific case of the middle wall, some of those particles are deleted in a certain stage of the simulations  for making the circular hole (this is discussed later in the text) and this number will depend on the open area under study. For ``frozen'' particles we mean that the resultant force acting on those particles are set zero.

Energies and lengths are given in terms of $\epsilon$ and $\sigma$, respectively. All the other quantities can be 
reduced to adimensional forms (from now on, just \emph{reduced units}) if written in terms of $\epsilon$, $\sigma$, $m$ and Boltzmann constant $k_B$. For example, pressure $P$ is given in terms of $\epsilon/\sigma^3$.  Time and temperature are written in terms of 
$\sqrt{m \sigma/\epsilon}$ and $\epsilon/k_B$, respectively. 

Initially 12480 particles were randomly placed in the Region 1 while the Region 2 was kept empty. Among those particles, 12000 are solvent, 240 are plus ions, and 240 are minus ions. As a result of external pressure applied in the piston, particles may eventually
permeate the middle wall through the circular pore. In order to avoid excessive pressure fluctuations in the fluid, the piston mass was set to be 1000 times the mass of solvent particles $m$. A viscous force proportional
to the piston velocity was used with the constant of proportionality  set as 100 in reduced units. 
Solute and solvent particles have the same mass.

Molecular dynamics simulations were performed with  LAMMPS package \cite{plimpton1995fast}
in the canonical ensemble using the Nos\'e-Hoover thermostat. 
The simulations were divided into equilibration and production stages. The equilibration stage was 
subdivided into two substages as follows. First, the solute-solvent system was evolved during 150,000 steps with initial temperature $\widetilde{T} = k_B T/\epsilon=2$ and final 
temperature $\widetilde{T}=1$ confined to Region 1 (the pore was kept blocked during the whole
equilibration stage in order to prevent particle permeation). Temperature linearly decreased  during this process. Secondly,  
the system was simulated for further 150,000 steps at $\widetilde{T} =1$ (still confined into Region 1), when configurational energy and pressure showed to be stable, i.e.,  fluctuate around average values. Following the equilibration stage, the production
stage consisted in (i) deleting atoms responsible for blocking the
pore located in the middle wall, (ii) applying an external pressure to the piston, and (iii) simulating the whole system for up to 4$\times10^6$ steps, depending on the quantity of interest under investigation. The timestep used in all runs was 0.001 in reduced units.

For the interaction between solvent particles we used the core-softened potential proposed by de Oliveira and collaborators \cite{de2006thermodynamic, de2006structural} , which reads 
$U_\mathrm{ss}/\epsilon = 4[(\sigma/r)^{12} -(\sigma/r)^6] + 5 \exp[-(r/\sigma-0.7)^2]$. Here $\sigma$ and $\epsilon$ are the well-known Lennard-Jones parameters while $r$ is the distance between pairs of solvent particles.
This model was previously used for mimicking water diffusion through carbon nanotubes presenting
good results compared to all-atoms simulations \cite{bordin2012diffusion}.

The ion-ion potential was described by the repulsive part of the Lennard-Jones potential plus a Coulombic interaction as $U_\mathrm{ii}/\epsilon^{\prime}=C z_1 z_2/ r+4[(\sigma^{\prime}/r)^{12}-(\sigma^{\prime}/r)^{6}]$. $z_1$ and $z_2$ can take values $+1$ or $-1$, corresponding to the valence of 
ions. Here, $\sigma^{\prime} = 3\sigma$ and $\epsilon^{\prime}=\epsilon$. $C=0.25\sigma$ is a constant associated to the strength of the Coulombic interaction and $r$ stands for the distance between ions. The cutoff distance for the Lennard-Jones and the Coulomb's potentials were 3.0 and 7.5, respectively, in units of $\sigma$. For the Coulomb's potential, interactions were
directly computed between pairs of particles for distances less than the cutoff. For distances 
greater than this value, Particle-Particle Particle-Mesh scheme was adopted \cite{PPPM}. Finally, the interaction between  solvent and ion particles was described by a Lennard-Jones potential,
$U_\mathrm{si}/\epsilon^{\prime\prime} = 4 [(\sigma^{\prime\prime}/r)^{12}-(\sigma^{\prime\prime}/r)^{6}]$ with $\epsilon^{\prime\prime} =\epsilon/2$ and $\sigma^{\prime\prime} = (\sigma+\sigma^{\prime})/2= 2\sigma$. This solvent-solute model was conceived inspired by the work of Luksic and coworkers \cite{lukvsivc2012structural}. 
The interaction between particles which compose the piston  and walls and those which compose the fluid (both solvent and solute ones) was considered as the repulsive repulsive part of the Lennard-Jones potential with energy strength $\epsilon$. The size of piston and walls particles were set to $2\sigma$.

\begin{figure}[ht]
\includegraphics[clip,scale=0.5]{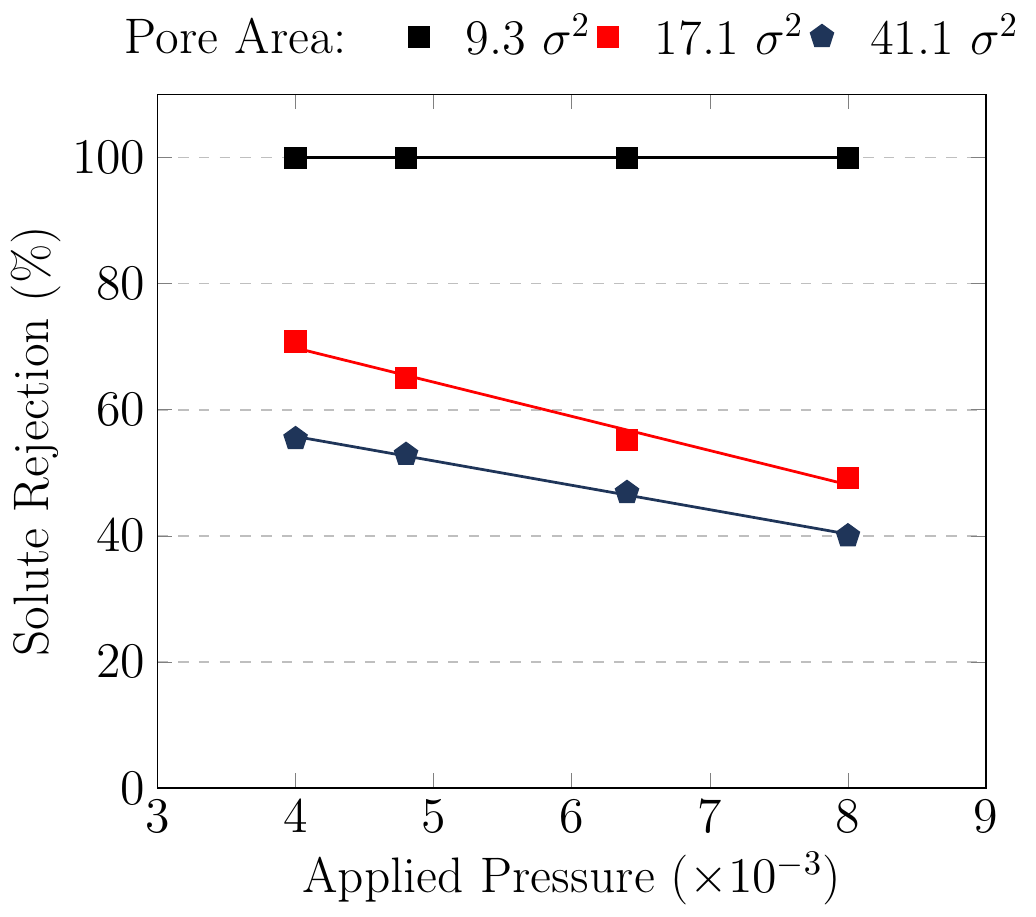}
\caption{Solute rejection percentage of a porous membrane as a function of the pressure applied
for different areas. Symbols are simulated data and lines are linear fittings through the data. \label{fig:rejection}}
\end{figure}
\begin{figure}[ht]
\includegraphics[clip,scale=0.5]{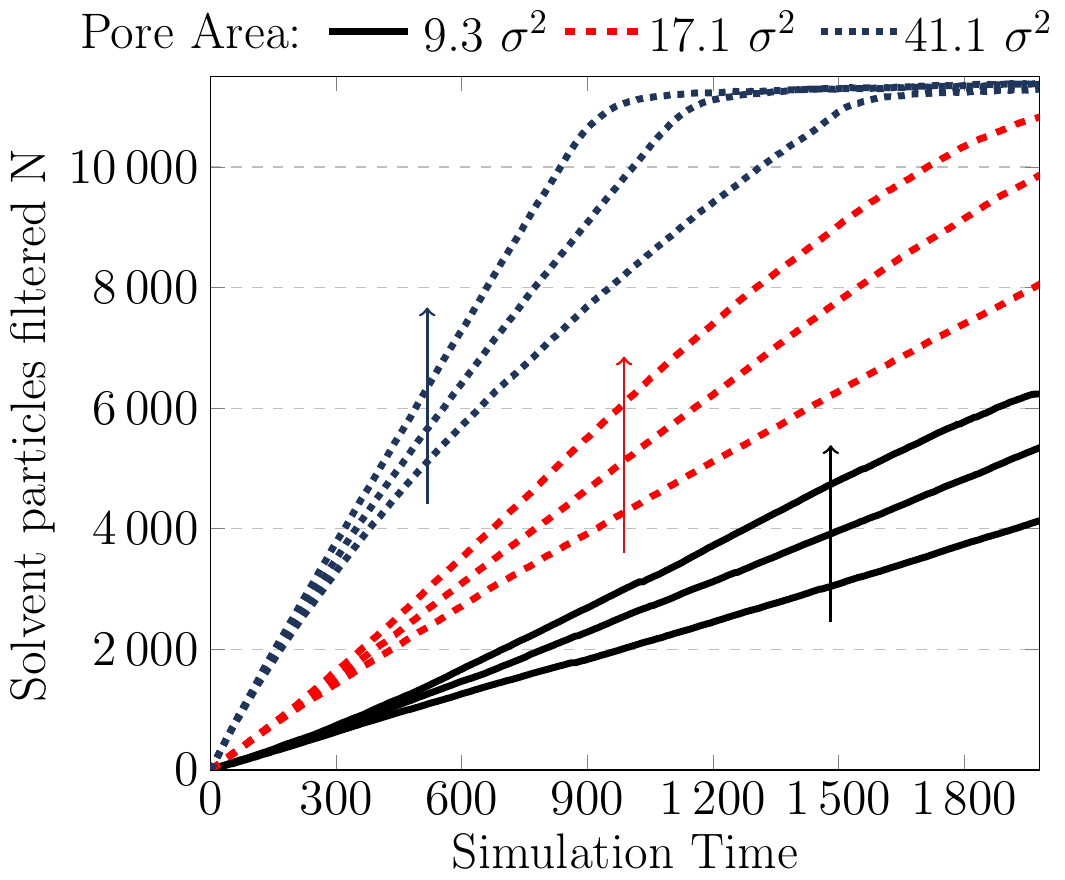}
\caption{Percentage of solvent particles that passed through the membrane (particles filtered) as a function of simulated time. Arrows indicate pressures 0.0010, 0.0014, and 0.0018 in reduced units from bottom to top. 
\label{fig:filter}}
\end{figure}

This simple, computationally cheap model was able to qualitatively reproduce results seen in Cohen-Tanugi and Grossman's work \cite{cohen2012water}. For example,  Fig. \ref{fig:rejection} presents the 
percentage of solute retained in the Region 1 as a function of external applied pressure while Fig. \ref{fig:filter} shows the number of solvent particles filtered over time for different pore sizes.

The results shown in  Figure \ref{fig:rejection} refer to the rejection parameter defined as 
$R=1-C_2/C1$ at the time in which half of the solvent molecules permeate the membrane and reach Region 2. $C_1$ and $C_2$ are the solute concentrations in  Region 1 and Region 2, respectively. Not surprisingly, for very small pore areas the rejection is 100\% independently the external pressure due to size exclusion, while for wider pores  it diminishes. For each pore size the rejection efficiency linearly decreases with external applied pressure.

 Results seen in Fig. \ref{fig:filter} show the dependence
of the number of permeate particles over time with both pore area and external applied pressure. For each pore
area we investigated pressures 1.0, 1.4, and 1.8
in units of $\times10^{-3}\epsilon/\sigma^3$ (from bottom to top). 
We see from this figure the linear dependence between $N$ and time for all pressure and pore
areas investigated. For higher pressures, the number of permeate particles rapidly saturate  due to the finitude 
of the system while for lower pressures longer times would be necessary for reaching such a saturation. 
Both solute rejection and filtered solvent particles behaviours agree with fully atomistic simulations 
found in the literature \cite{cohen2012water}.

\begin{figure}
\includegraphics[clip,scale=0.6]{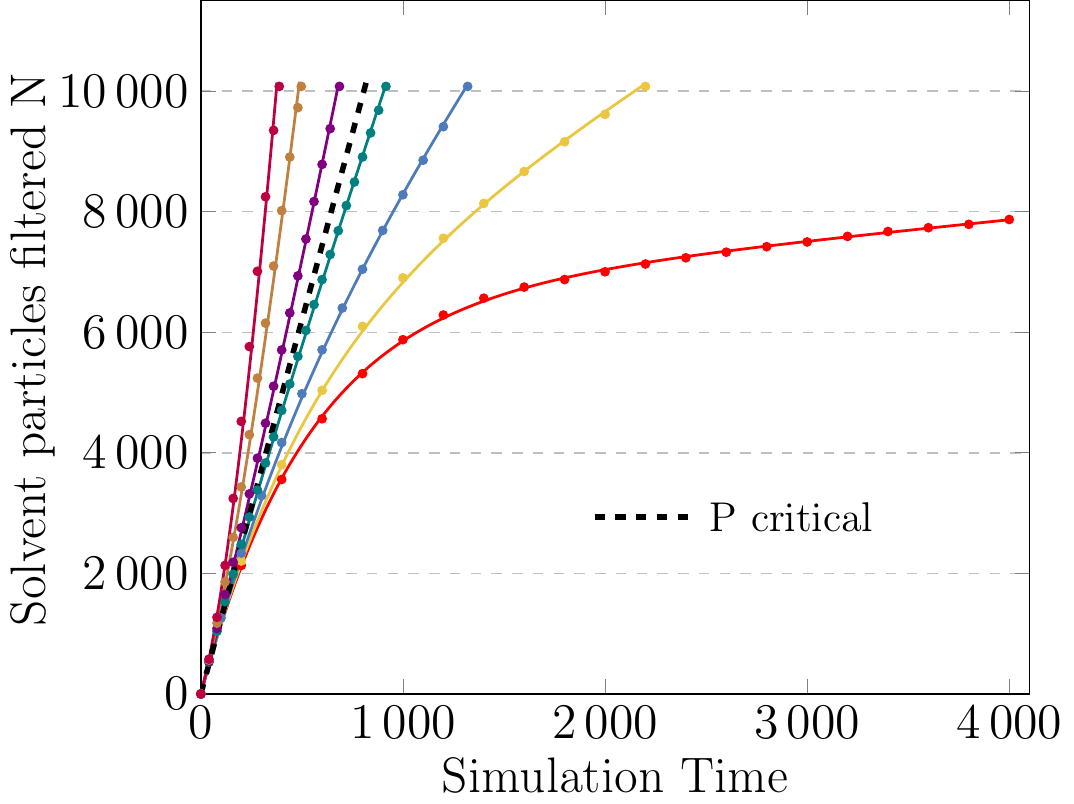}
\caption{Same as Fig. \ref{fig:filter} but for a single pore area (41.1$\sigma^2$) and
a wide range of pressures. Here, from bottom to top, they are 0.0002, 0.0006, 
0.0010, 0.0016, 0.0024,  0.0040, and 0.0064 in reduced units. 
Symbols are data from simulations and curves are fittings using our theoretical model Eq. (\ref{eq:solution}).
The parameters used in the fittings were $\lambda=0.0018$ and $\phi A$ = 12.5, both in reduced units, for all curves. For $\mu P$, we have used the values 0.3, 2.1, 4.8, 9.6, 17.8, 36.4, and 63.1 in reduced units. The dashed line is the theoretical  critical pressure, obtained by using $\phi A = \mu P =$ 12.5 in the Eq. (\ref{eq:solution}).\label{fig:fit}}
\end{figure}

Due to the simplicity of our model, we were able to investigate not only bigger systems (compared with typical system sizes seen in literature), as well as broader space of parameters. 
 For example, in Fig. \ref{fig:fit} we show results for the number of permeate solvent particles as a function
 of time for the  pore area of 41.1$\sigma^2$. Differently of what is shown in Fig. \ref {fig:filter}, here we show pressures  ranging from 0.0002 up to 0.0064 (see figure caption for details). In this figure, symbols are simulated data and curves are fittings based on our theoretical approach, which will be presented soon in this Letter.
 It is clear that depending on the chosen pressure 
 the dependence between $N$ and $t$ is far from linear (as is seen in Fig. \ref{fig:filter}), demonstrating that   $N(t)$  has a rich relationship with both pressure and pore area under investigation. For explaining the
  various regimes seen in Fig. \ref{fig:fit} we developed a theoretical model as follows. 
 
Let $N,$ $A,$ and $P$ be the number of permeated particles, hole area and
external pressure upon the system. The rate in which particles come across the hole of area $A$, $dN/dt$, must  be proportional to its area. On the other hand, $dN/dt$ must decrease with the number of permeate particles. Thus we will assume that $dN/dt \propto -N$. Also we must include the dependence of $dN/dt$ with the external pressure. If $\eta$ is the fluid viscosity, such a dependence can be written as $dN/dt\propto Pt/\eta$. This quantity is adimensional and it is proportional to the relative displacement  between different layers of fluid under shear stress (this will be discussed in details below).  Considering the model described above, the differential equation which governs the flux of particles to the permeate region is given by

\begin{equation}
\frac{dN}{dt}=\phi A-\lambda N+\gamma \frac{Pt}{\eta},\label{eq:model}
\end{equation}

\noindent where $\phi,$ $\lambda,$ and $\gamma$ are proportionality constants. 
$\phi$ has the physical meaning of flux of particles (number of particles per time per pore area) whereas 
$\lambda$ give us the information of how difficult is for a particle to pass from Region 1 to Region 2
once there are already $N$ particles in the Region 2. The interpretation of $\gamma$ is a little more
involving as follows. We argument that due to the circular porous particles localised in the central regions 
of the simulation box will diffuse faster than those in the borders. We mentally represent those group of atoms 
as a cylinder surrounded by a box of particles. 
Particles in the middle of the cylinder will diffuse more than those localised in its borders because of the shear stress due to the contact with the even less mobile particles which compose the  surrounding box. In a general case, for two layers of fluid with relative constant velocity $v$, area $\cal{A}$ and distant $y$ apart, the viscosity $\eta$ is given by
$F/{\cal A} = \eta v/y$, with $F$ being the force responsible for the relative velocity between layers. Regarding our model, $F$ is  proportional to the external pressure imposed by the piston upon the system and we can write
$\varphi  P = \eta (z/t)/y$. Here $z$ is the displacement during the time $t$
 and $\varphi$ is the ratio between the piston area and the moving layer area under shear stress. From this equation we conclude that for a given pore area $Pt/\eta\propto z$. Thus $\gamma$ in the 
Eq. (\ref{eq:model}) gives the number of particles per time crossing the pore due to the 
displacement of a group of particles localised in the center of the simulation box because of the 
piston action.

Back to the Eq. (\ref{eq:model}), it is a linear differential equation whose solution is given by

\begin{equation}
N(t)=\frac{1}{\lambda}\left(\phi A- \mu P\right)\left(1-e^{-\lambda t}\right)+\mu Pt,\label{eq:solution}
\end{equation}

\noindent with  $\mu = \gamma/\lambda\eta$ and we have used the initial condition $N=0$ at $t=0$.
One important point regarding Eq. (\ref{eq:solution}) is that if
$P=P_{c}=A \phi/\mu$ the first term on the right side vanishes
and the relation between $N(t)$ and $t$ becomes linear having $\mu P$
as the angular coefficient (see the dashed line in Fig. \ref{fig:fit}). 
We will call such a pressure $P_{c}$ as critical
pressure. Pressures around such a value will give virtually linear behaviours, as seen in Fig. \ref{fig:filter}.
For pressures below and above $A \phi/\mu$ the
terms subcritical and supercritical, respectively, seem to be appropriated. 
The critical, subcritical, and supercritical behaviours become
clearer if we analyse the dependence of $N$ with time for short and long times separately. For example, 
if $\lambda t\ll1$, we end up with $N(t)\approx\phi A t$. This means that for shorter times 
the dominant term is related to the pore size and we expect that only particles close to the 
pore will randomly permeate due to essentially brownian movements.  The curve 
$N$ vs. $t$ must be insensitive to  external pressures and all pressure-curves must 
have virtually the same inclination $\phi A$. This is exactly  what we have found
through simulations as one can see in Fig. \ref{fig:fit} for very short times. 
In the another extremum, if $\lambda t\gg1$, we can
write $N(t)\approx b+at$, where $b=(\phi A - \mu P)/\lambda$ and $a=\mu P$. Again, we
 finish with a linear dependence between $N$ and $t$, with angular coefficient given by $\mu P$.
 For longer times, the flux of particles are mainly due to the piston action  with low 
 contribution from random insertions. For bigger $t$,
 we will have different curvatures depending on the  balance between pore area and external pressure
 as follows. In the initial times, all pressure-curves will collapse obeying the already mentioned $\phi A t$ dependence.
 As time increases, the term with $[1-\exp(-\lambda t)]$ becomes increasingly important and it bends the curves
to smoothly assume different curvatures from $\phi A$. 
This rich scenario is shown in Fig. \ref{fig:fit}, in which the simulation data (symbols) are 
beautifully fitted (curves) with our model Eq. (\ref{eq:solution}).



To conclude this Letter, we believe our results can contribute to understand 
the underlaying physics of  water desalination through nanostructures processes. We  present
both a simple, computationally cheap model based on the so-called core-softened potentials
and a theoretical approach which reasonable explains the flux of solvent
through a porous membrane for a wide variety of membrane porosity and external pressures. 
While our simulations have shown to be fairly robust against fully atomistic simulations, presenting 
good qualitative  results, the theoretical model clarified the behaviour of filtered particles 
with time for a wide range of parameters. This is important since it is usual to extrapolate the results 
obtained through simulations -- generally for high pressures and small pore areas --  to low pressures and
highly porous membranes.

The authors thank  the Brazilian science agencies CNPq, CAPES, and FAPEMIG for financial support.

\end{document}